\documentclass[aps,pre,twocolumn,amsmath,amssymb,superscriptaddress,reprint,longbibliography]{revtex4-1}
\usepackage{amsfonts,amsmath,amssymb,bm}
\usepackage{graphicx,graphics,float}
\usepackage{bm}
\usepackage{amssymb}
\usepackage{colordvi}
\usepackage{graphicx}
\usepackage{color}
\usepackage[colorlinks=true,linkcolor=blue,citecolor=blue,urlcolor=blue]{hyperref}%
\usepackage{hyperref}
\usepackage{comment}
\usepackage{harpoon}
\usepackage{braket}
\newcounter{supplementalequation}

\begin{document}

\title{Impact of quantum coherence on inelastic thermoelectric devices: From diode to transistor}

\author{Bei Cao}
\address{Jiangsu Key Laboratory of Micro and Nano Heat Fluid Flow Technology and Energy Application, School of Physical Science and Technology, Suzhou University of Science and Technology, Suzhou, 215009, China}

\author{Chongze Han }
\address{Jiangsu Key Laboratory of Micro and Nano Heat Fluid Flow Technology and Energy Application, School of Physical Science and Technology, Suzhou University of Science and Technology, Suzhou, 215009, China}

\author{Xiang Hao}
\address{Jiangsu Key Laboratory of Micro and Nano Heat Fluid Flow Technology and Energy Application, School of Physical Science and Technology, Suzhou University of Science and Technology, Suzhou, 215009, China}

\author{Chen Wang }\email{wangchen@zjnu.cn}
\address{Department of Physics, Zhejiang Normal University, Jinhua, Zhejiang 321004, China}

\author{Jincheng Lu}\email{jinchenglu@usts.edu.cn}
\address{Jiangsu Key Laboratory of Micro and Nano Heat Fluid Flow Technology and Energy Application, School of Physical Science and Technology, Suzhou University of Science and Technology, Suzhou, 215009, China}

\date{\today}
\begin{abstract}

We present a study on inelastic thermoelectric devices, wherein charge currents and electronic and phononic heat currents are intricately interconnected. The employment of double quantum dots in conjunction with a phonon reservoir positions them as promising candidates for quantum thermoelectric diodes and transistors. Within this study, we illustrate that quantum coherence  yields significant charge and Seebeck rectification effects. It's worth noting that while the thermal transistor effect is observable in the linear response regime, especially when phonon-assisted inelastic processes dominate the transport, quantum coherence does not enhance thermal amplification. Our work may provide valuable insights for the optimization of inelastic thermoelectric devices. 
\end{abstract}

\maketitle

{\emph {Introduction.}} Phonon-based thermoelectricity has attracted significant research attention, owing to their relevance in both fundamental physics and cutting-edge energy applications~\cite{RenRMP,Thier2015,BenentiPR}. Phonon-thermoelectric transport stands at the intersection of mesoscopic physics and open quantum systems, exploring topics such as thermoelectric energy conversion in phonon-thermoelectric devices~\cite{Jiang2012,Jiang2017} and quantum dot (QD) circuit-quantum-electrodynamics~\cite{WangPRE,LuEntropy,ChinPhysLett.40.090501}. 
Recently, various quantum phenomena based on QDs subjected to voltage bias and temperature gradients have been explored, revealing stressing the influence of quantum coherence  on electron and phonon transport~\cite{WuPRL17,CamatiPRA19,SegalPRE21,KwonPRA23,BeheraPRB23,ChinPhysLett.40.050501,ChinPhysLett.40.054401}. 
Coherence offers a promising viewpoint for investigating non-equilibrium open quantum systems, providing valuable insights for both theoretical foundations and practical applications~\cite{EngelNature,Vaziri10,SanchezPRB22,IvanderNJP22,Yuan2023,ChinPhysLett.40.117301}. However, the exploration of phonon-thermoelectric devices with quantum coherence, e.g.,  diodes and transistors, is still on demand.

Inelastic thermoelectricity recently emerges as a rising front field,  which opens up great potential for high-performance thermoelectric materials~\cite{MyReview,OraPRB2010,MyPRBdemon, Xi21CPL,PhysRevB.108.085430,PhysRevB.107.L241405}. In a multiterminal setup, such as a three-terminal system, electronic and phononic heat currents are nonlinearly coupled due to inelastic electron-phonon interactions. These interactions involve at least three reservoirs, which is why we refer to the process as ``inelastic transport." This term highlights the involvement of particles from different terminals, resulting in energy exchange facilitated by phonons. The control and separation of heat and electrical currents on mesoscopic scales have gained fundamental importance~\cite{PekolaPRM21, LvPRB16, Li21NRM, NianPRB23}. In particular, phonon-thermoelectric diode and transistor can be realized even in linear response regime, in absence of negative differential thermal conductance~\cite{Jiangtransistors}. 
Besides, inelastic thermoelectricity greatly broad the 
the working bound of quantum thermal devices~\cite{MyPRBtransistor}.

In this paper, we delve into the realm of inelastic thermoelectric transport facilitated by phonons. We demonstrate that quantum coherence provides a promising avenue for realizing quantum thermoelectric devices. 
by employing the quantum master equation with  full counting statistics approach, we showcase how the electron-phonon interactions induced nonlinearity allows for significant charge and Seebeck rectification effect by  tuning  the QD's energy levels. Furthermore, we illustrate that a three-terminal setup can exhibit thermal transistor effects even within the linear transport regime, whereas quantum coherence is detrimental to  heat amplification effect. Our work establishes a robust platform offering unprecedented control over both charge and heat. Throughout this paper, we set natural units $\hbar=k_B=e\equiv 1$.

\begin{figure}[htb]
\begin{center}
\centering\includegraphics[width=8.5cm]{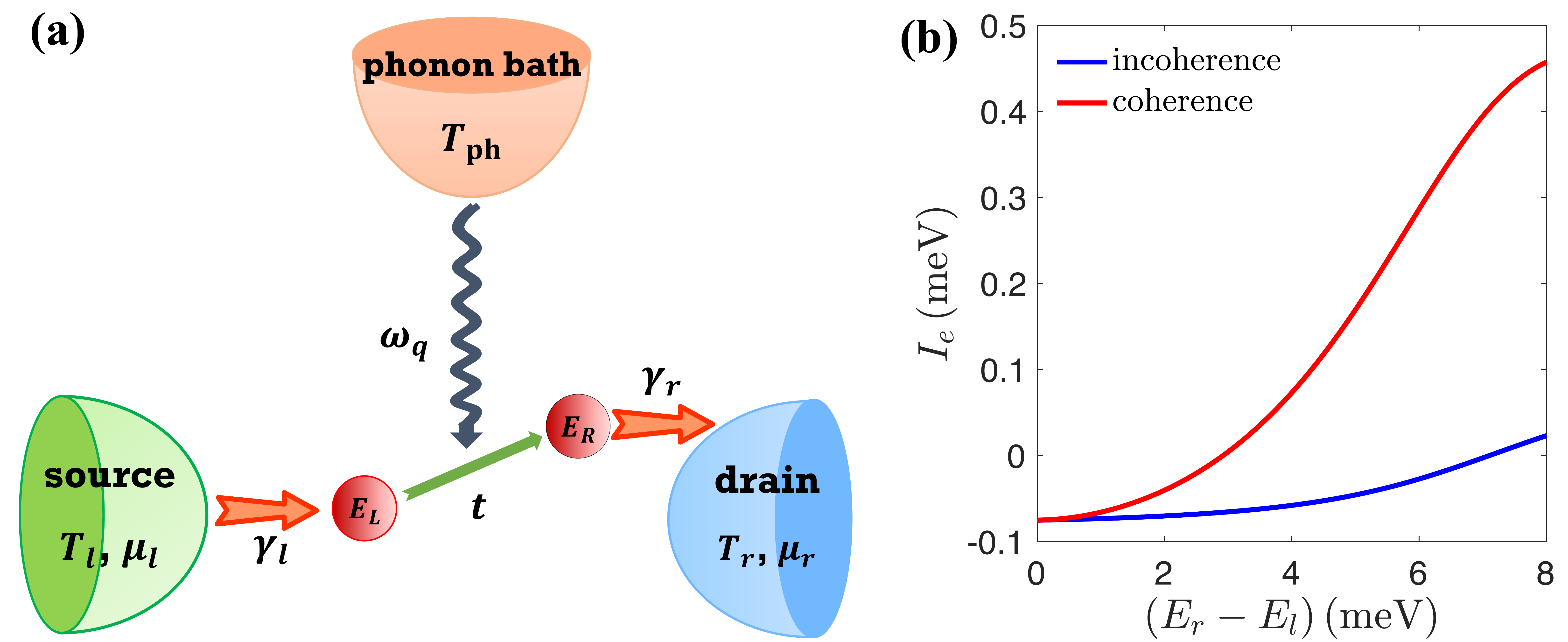}
\caption{(a) Illustration of the three-terminal inelastic thermoelectric device. An electron left the source into the left QD with energy $E_l$ hops to the right QD with a different energy $E_r$ assisted by a phonon from the phonon reservoir (with temperature $T_{\rm ph}$). The temperatures and chemical potentials of two electric reservoirs are $T_{L(R)}$ and $\mu_{L(R)}$, respectively. The temperature of phonon reservoir is $T_{\rm ph}$. $t$ is the tunneling between the two QDs, $\gamma_{l}$ ($\gamma_{r}$) is the coupling between the dots and the left (right) reservoir, $\lambda_q$ is the strength of electron-phonon interaction.  (b) The charge current $I_e$ as a function of level detuning $E_r-E_l$. The other parameters are given by $\mu=6\, {\rm meV}$, $\Delta\mu = 2 \, {\rm meV}$,  $E_l=0$, $\Gamma_l=\Gamma_r=\Gamma_{\rm ph}=6 \, {\rm meV}$, $t=8 \, {\rm meV}$, and $k_BT_l=k_BT_r=k_BT_{\rm ph}=10\, {\rm meV}$. }\label{fig:3Tsystem}
\end{center}
\end{figure}

{\emph {Model and method}.} In our setup (see Fig.~\ref{fig:3Tsystem}(a)), a double QD system is coupled to a phonon reservoir ($\rm ph$) and exchanging energy with two individual electronic reservoirs marked as $L$ and $R$. 
The Hamiltonian reads~\cite{Jiang2012}
\begin{equation}
\hat H = \hat H_{\rm DQD} + \hat H_{\rm e-ph} + \hat H_{\rm lead} + \hat H_{\rm tun} + \hat H_{\rm ph}.
\end{equation}
$\hat H_{\rm DQD} = \sum_{i=l,r} E_i \hat d_i^\dagger \hat d_i +  t(\hat d_l^\dagger \hat d_r + {\rm H.c.})$ describes the QDs with their energies $E_i$, tunneling $t$, 
and electron creating(annihilating) operator 
$\hat d_i^\dagger (\hat d_i)$ in the $i$-th dot. 
$\hat H_{\rm e-ph} = \sum_{q}\lambda_q \hat d_l^\dagger \hat d_r (\hat a_q + \hat a^\dagger_q) + {\rm H.c.}$ accounts for electron-phonon interactions. 
$ \hat H_{\rm ph} = \sum_q\omega_{q}\hat a^\dagger_q \hat a_q$ represents the phonon reservoir.
$\hat H_{\rm lead} = \sum_{j=L,R}\sum_{k} E_{jk} \hat d_{jk}^\dagger \hat d_{jk}$ characterizes the electronic reservoirs' energy levels. 
$\hat H_{\rm tun} = \sum_k \gamma_{lk} \hat d_l^\dagger \hat d_{Lk} + \sum_k \gamma_{rk} \hat d_r^\dagger \hat d_{rk} +  {\rm H.c.}$ involves the tunneling between  QDs and the reservoirs.

To analyze the thermoelectric device within the eigenbasis of the double QDs, we initiate the diagonalization of $\hat H_{\rm DQD}$ as follows:
\begin{equation}
\hat H_{\rm DQD} = E_D\hat D^\dagger\hat D + E_d\hat d^\dagger\hat d,
\end{equation}
$E_D = \frac{E_r +E_l }{2}+\sqrt{\frac{(E_r -E_l )^2}{4}+\Delta^2}$ and $E_d = \frac{E_r +E_l }{2}-\sqrt{\frac{(E_r -E_l )^2}{4}+\Delta^2}$
represent the eigenenergies, respectively.
The new sets of Fermion operators are defined as
$\hat D = \sin\theta \hat d_l + \cos\theta \hat d_r$ and $\hat d = \cos\theta \hat d_l - \sin\theta \hat d_r$~\cite{MyPRBdiode}, where $\theta$ denotes $\arctan\left(\frac{2\Delta}{E_r -E_l }\right)/2$.
Consequently, the electron-phonon and dot-reservoir tunneling terms are reformulated as $\hat H_{\rm e-ph} = \sum_q\lambda_q [\sin(2\theta)(\hat D^\dagger\hat D-\hat d^\dagger\hat d) + \cos(2\theta)(d^\dagger D+D^\dagger d)](\hat a^\dagger_q + \hat a_q)$ and $ \hat H_{\rm tun} = \sum_k [\gamma_{lk} (\sin\theta\hat D^\dagger + \cos\theta\hat d^\dagger) \hat d_{Lk},\gamma_{rk} (\cos\theta\hat D^\dagger - \sin\theta\hat d^\dagger) \hat d_{rk}] + {\rm H.c.}$.
It can be inferred from the term $\hat{H}_{\rm e-ph}$ that within the eigenbasis of $\hat{H}_{\rm DQD}$, both dephasing and damping processes are included, which may cooperatively give rise to the steady-state coherence.

Based on two-time measurement protocol~\cite{CampisiRMP},
we apply the full counting statistics to obtain the particle and energy flows out of electronic reservoirs and heat current out of phonon reservoir by including 
$\mathbf{\chi}=\{\lambda_p$, $\lambda_E$, $\lambda_{\rm ph}\}$, respectively.  
Consequently, the counting-field-dependent total Hamiltonian is described as    
\begin{equation}
H_{-\mathbf{\chi}/2} 
= H_{\rm DQD} + H_{\rm ph} + H_{\rm lead} + V_{-\mathbf{\chi}/2}, 
\end{equation}
with $V_{-\mathbf{\chi}/2}$ specified as         
\begin{equation}
\begin{aligned}
V_{-\mathbf{\chi}/2} &= \sum_q\lambda_q [\sin(2\theta)(\hat D^\dagger\hat D-\hat d^\dagger\hat d) \\
&+ \cos(2\theta)(d^\dagger D+D^\dagger d)](e^{i\frac{\lambda_{\rm ph}}{2}\omega_q}{\hat a}_q +  {\rm H.c.}) \\
&+ \sum_k ([\gamma_{lk} (\sin\theta\hat D^\dagger + \cos\theta\hat d^\dagger) \hat d_{Lk}  \\
&+  \gamma_{rk}e^{-i\frac{\lambda_p}{2}-i\frac{\lambda_E}{2}E_{rk}} (\cos\theta\hat D^\dagger - \sin\theta\hat d^\dagger) \hat d_{rk}] +  {\rm H.c.}).  
\end{aligned}
\end{equation}
We assume the electron-phonon coupling and dot-reservoir tunnelings are weak. Based on the Born-Markov approximation,
we perturb $V_{-\mathbf{\chi}/2}$ to obtain the quantum master equation as
\begin{equation}
\begin{aligned}
&\frac{\partial}{\partial t} \rho_S(\mathbf{\chi},t)
= i[\rho_S(\mathbf{\chi},t),H_{\rm DQD}] \\
&- \int_0^\infty d\tau {\rm  Tr}_B\{[
[V_{-\mathbf{\chi}/2},[V_{-\mathbf{\chi}/2}(-\tau),
\rho_S(\mathbf{\chi},t){\otimes}\rho_B]_{\mathbf{\chi}}
]_{\mathbf{\chi}}\}, 
\label{eq:rhoSV}
\end{aligned}
\end{equation}
where $\rho_S(\mathbf{\chi},t)$ denotes the reduced density operator of {central double QDs system} with counting parameters, i.e.
$\rho_S(\mathbf{\chi},t)
=\textrm{Tr}_B\{\rho^T_{\mathbf{\chi}}(t)\}$,
with $\rho^T_{\mathbf{\chi}}(t)$ the full density operator of the whole inelastic thermoelectric device,
the commutating relation denotes
$[\hat{A}_{\mathbf{\chi}},\hat{B}_{\mathbf{\chi}}]_{\mathbf{\chi}}
=\hat{A}_{\mathbf{\chi}}\hat{B}_{\mathbf{\chi}}
-\hat{B}_{\mathbf{\chi}}\hat{A}_{-\mathbf{\chi}}$,
and the equilibrium distribution of reservoirs is specified as
$\rho_B=\rho_l{\otimes}\rho_r{\otimes}\rho_{\rm ph}$,
with $\rho_i$ is the density of matrix of the $i$-reservoir ($i=L,R,\rm ph$).   
If we reorganize $\rho_S(\mathbf{\chi},t)$ in the vector form
$| {\mathbf P}(\mathbf{\chi},t) \rangle$ = $[\langle 0 |\rho_S (\mathbf{\chi},t) | 0 \rangle$; $\langle D |\rho_S (\mathbf{\chi},t) | D \rangle$; $\langle d |\rho_S (\mathbf{\chi},t) | d \rangle$; $\langle D |\rho_S (\mathbf{\chi},t) | d \rangle $; $\langle d |\rho_S (\mathbf{\chi},t) | D \rangle ]$,
the quantum master equation is reexpressed as
\begin{equation}
\begin{aligned}
\dfrac{d | {\mathbf P}({\mathbf{\chi},t}) \rangle} {dt}= {\mathbf H}(\mathbf{\chi})| {\mathbf P}(\mathbf{\chi},t)\rangle, 
\end{aligned}
\end{equation}
where  $ {\mathbf H}(\mathbf{\chi})$ is the evolution matrix.  We note that the nonzero off-diagonal elements of the density matrix, i.e., $\langle D |\rho_S (\mathbf{\chi},t) | d \rangle $ and $\langle d |\rho_S (\mathbf{\chi},t) | D \rangle $,
are the signature of quantum coherence.
Conversely, if we want to neglect the quantum coherence effect, we consider Secular approximation to disregard the off-diagonal elements (the details see supplementary materials~\cite{Supp}).

The charge, energy, and phonon currents flowing from the reservoir into the system are expressed as
\begin{equation}
I_e^r = \frac{\partial{\mathcal G}}{\partial (i\lambda_e)}|_{\chi=0}, \, I_E^r  =  \frac{\partial{\mathcal G}}{\partial (i\lambda_E)}|_{\lambda=0}, I_Q^{\rm ph}  =  \frac{\partial{\mathcal G}}{\partial (i\lambda_{\rm ph})}|_{\chi=0}.
\end{equation}
respectively, where the ${\mathcal G}=\lim_{t\rightarrow \infty}\frac{1}{t} \ln \langle I | {\mathbf P}({\mathbf{\chi},t}) \rangle$ is the cumulant generating function,
which can be alternatively expressed as the
minimal real part of the eigenvalues of the evolution matrix ${\mathbf H}(\chi)$~\cite{wangpump}. The electronic heat current extracted from the right reservoir is defined as  $I_Q^r  = I_E^r -\mu_rI_e^r$~\cite{YamamotoPRE15}. Similarly, the particle current $I_e^l$ and energy current $I_E^l$ flowing from the left ($L$) reservoir into the central system can also be obtained by introducing the counting functions.  Charge conservation implies that $I_e^l + I_e^r=0$, while the energy conservation requires that $I_E^l + I_E^r +I_Q^{\rm ph}=0$~\cite{JiangCRP}. 

Before delving into detailed discussions, it is essential to note that the methods employed in our study rely on weak system-reservoir couplings, denoted as $\gamma_i$ ($i=l, r, \mathrm{ph}$). In a prior work~\cite{Lu23Coh}, we showcased the feasibility of periodically driven thermoelectric engines in a three-terminal double QDs system. Notably, we observed a substantial enhancement in both output work and efficiency due to the presence of quantum coherence. In Figure \ref{fig:3Tsystem}(b), with a fixed chemical potential difference of $\Delta\mu=2\, \mathrm{meV}$, we present the net charge current $I_e \equiv \frac{1}{2}(I_e^r-I_e^l)$  as a function of the QD energy difference $E_r - E_l$ for both coherent and incoherent transport. It is evident that quantum coherence, characterized by nonzero off-diagonal elements in the system's density matrix, leads to a notable improvement in the charge current when compared to the scenarios in absence of quantum coherence.

\begin{figure}[htb]
\begin{center}
\centering\includegraphics[width=8.5cm]{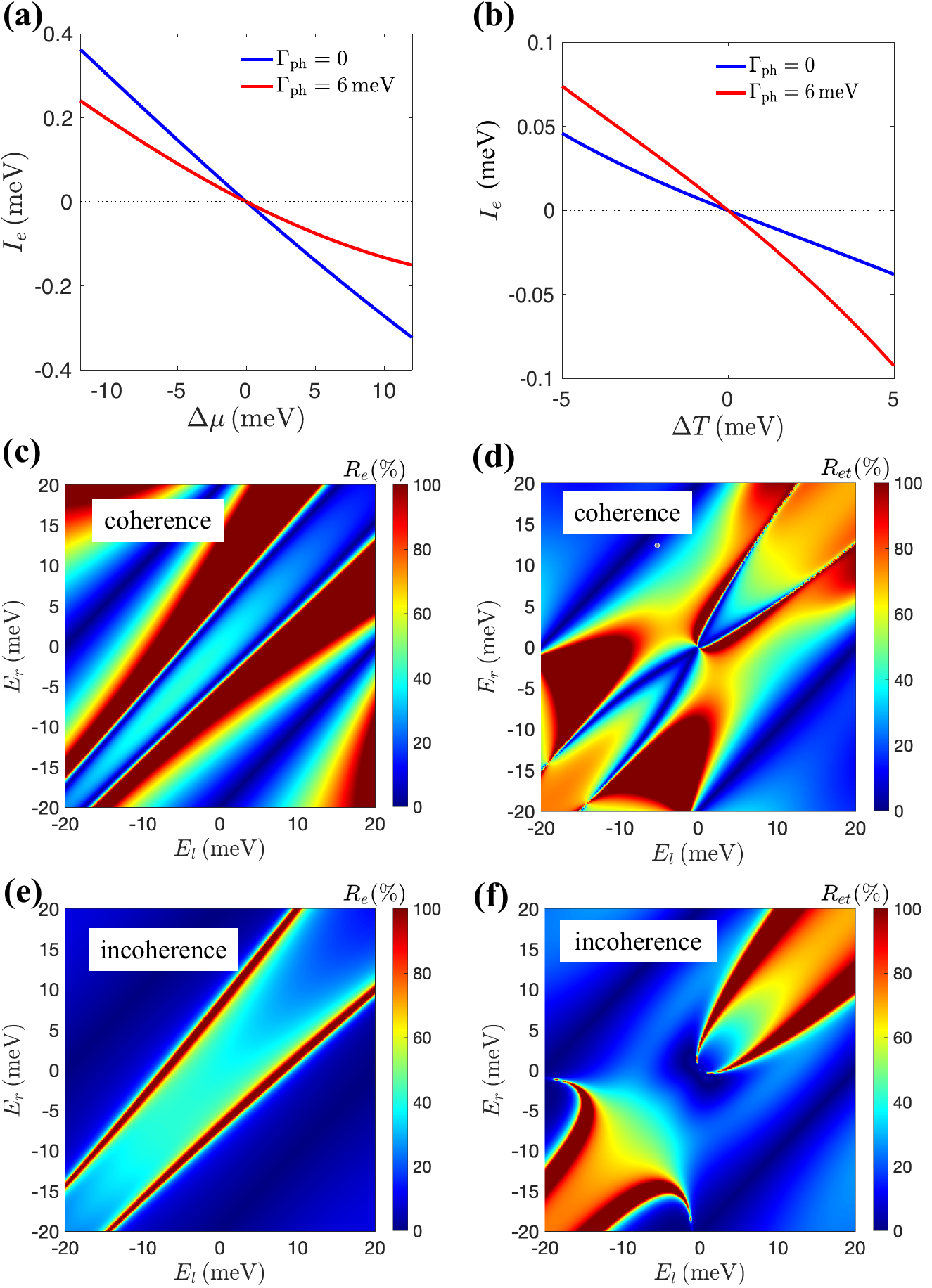}
\caption{Charge and Seebeck-rectification effects. (a) Charge current $I_e$ as a function of $\Delta\mu$ for different $\Gamma_{\rm ph}$. The parameters are $\mu=0$, $E_l=5\, {\rm meV}$, $E_r=7\, {\rm meV}$, $k_BT_{\rm ph}=10\, {\rm meV}$. (b) Charge current $I_e$ as a function of $\Delta T$ for different $\Gamma_{\rm ph}$, where $\Delta\mu=0$, $E_l=5\, {\rm meV}$, $E_r=7\, {\rm meV}$. The charge rectification $R_e$ as functions of QD energies $E_l$ and $E_r$ for (c) coherent transport and (e) incoherent transport, where $\Delta\mu=10\, {\rm meV}$ and $k_BT_{\rm ph}=10\, {\rm meV}$. The Seebeck-rectification $R_{et}$ as functions of QD energies $E_l$ and $E_r$ for (d) coherent transport and (f) incoherent transport, where $\Delta\mu=0$ and $k_BT_{\rm ph}=15\, {\rm meV}$. The other parameters are given by $\Gamma_l=\Gamma_r=6 \, {\rm meV}$, $t=8 \, {\rm meV}$, $k_BT_l=k_BT_r=10\, {\rm meV}$. }
\label{fig:diode}
\end{center}
\end{figure}

{\emph{ Diode effect}. }In this section, we will illustrate that the three-terminal configuration effectively behaves as the diode~\cite{ZhangPRB10,RenPRB13,ZhangPRE18,ZhangPhysicaA,WangPRE19,DazNJP21,TesserNJP22,KhandelwalPRR23}. Our system, renowned for its combined thermal and electrical transport capabilities, offers the potential for unconventional rectification effect. Notably, it can induce charge rectification owing to a temperature difference, in addition to  the conventional charge rectification driven by voltage bias~\cite{Jiangtransistors,MyPRBdiode}. 

In a general nonequilibrium setup,
the charge rectification effects can be characterized as
\begin{equation}
\begin{aligned}
R_e=\frac{|I_e(\Delta\mu)+I_e(-\Delta\mu)|}{|I_e(\Delta\mu)|+|I_e(-\Delta\mu)|},
\end{aligned}
\end{equation}
which is induced by chemical potential.
And such rectification induced by the temperature gradient, termed as Seebeck rectification, is expressed as~\cite{Jiangtransistors}
\begin{equation}
\begin{aligned}
R_{et}=\frac{|I_e(\Delta T)+I_e(-\Delta T)|}{| I_e(\Delta T)| + | I_e(-\Delta T)|}. 
\end{aligned}
\end{equation}

Here, we demonstrate both charge and Seebeck-rectification effects in Fig.~\ref{fig:diode} based on the three-terminal QDs  system. 
From Figs.~\ref{fig:diode}(a) and \ref{fig:diode}(b), we  find that the elastic heat and charge currents (i.e.,$\Gamma_{\rm ph}=0$) are almost symmetrical,
whereas the inelastic currents i.e., $\Gamma_{\rm ph}\ne 0$) become dramatic asymmetric. 
Hence, the microscopic inelastic  electron-phonon scattering 
generally enhances the asymmetry of currents~\cite{Jiangtransistors}.

Furthermore, we can make a comparative analysis between the charge rectification and Seebeck rectification effects in coherent and incoherent scenarios. 
Specifically, We  study the influence of the QDs' energies $E_l$ and $E_r$ on rectification effects with and without quantum coherence. 
As depicted in Figs.\ref{fig:diode}(c)-\ref{fig:diode}(e) and Figs.\ref{fig:diode}(d)-\ref{fig:diode}(f), it becomes evident that both charge ($R_e$) and Seebeck ($R_{et}$) rectification effects in thermoelectric devices, when considering quantum coherence, is significantly more pronounced than the one with incoherence. Hence, we can conclude that quantum coherence not only amplifies the current magnitudes but also intensifies the asymmetry of inelastic currents, thereby enhancing the rectification effects.

\begin{figure}[htb]
\begin{center}
\centering\includegraphics[width=8.5cm]{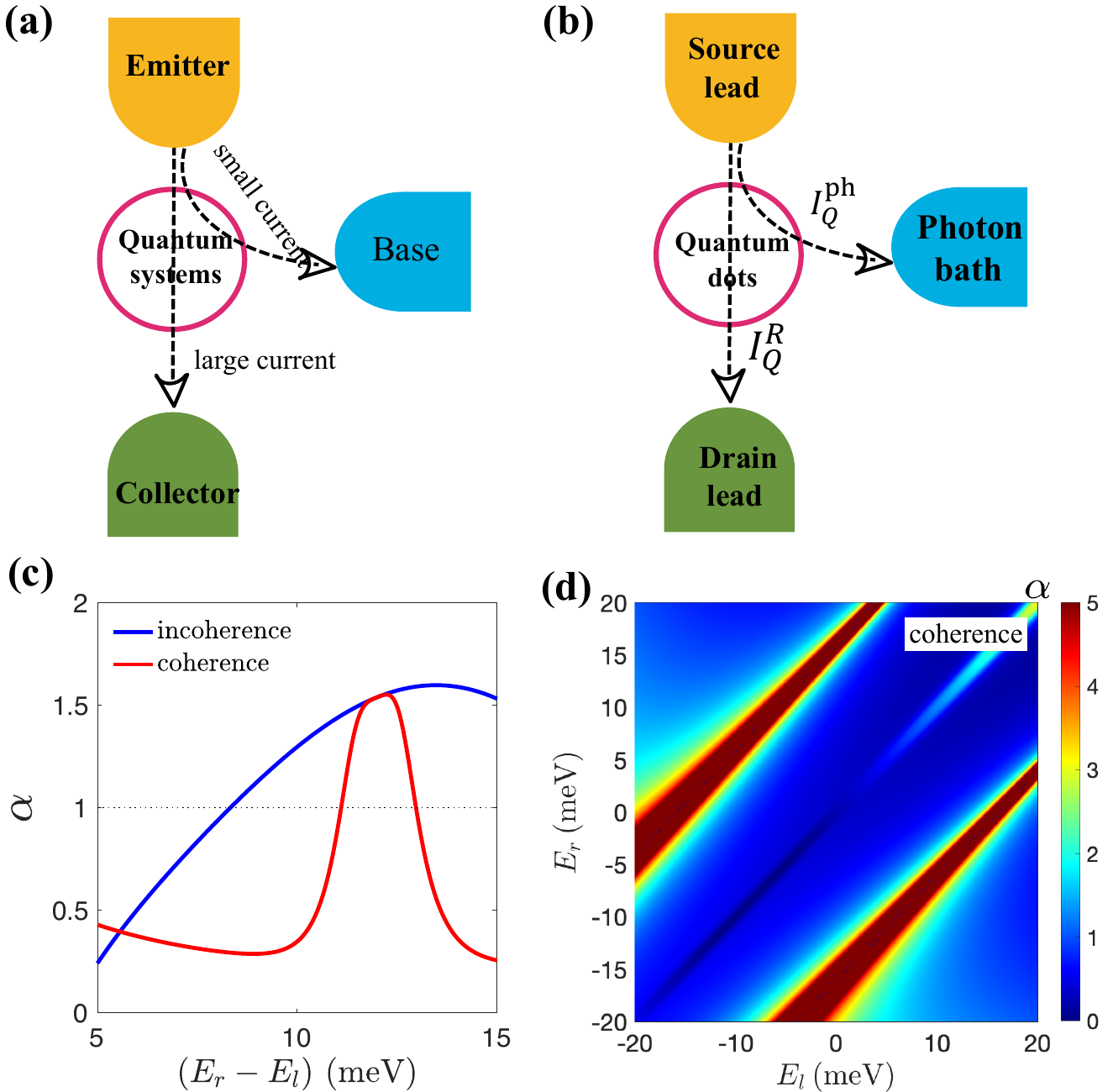}
\caption{(a) Schematic outlines the functioning of a conventional transistor. A small charge current originating from the emitter to the base governs a significantly larger charge current between the emitter and collector. The ratio between these two charge currents is the defining characteristic of the transistor effect (b) Scheme of a three-terminal quantum-dot system that can serve as a
phonon-thermoelectric transistor. Here, a modest heat current travels from the source lead to the phonon reservoir ($I_Q^{\rm ph}$), and it exerts control over a substantial heat current flowing from the source lead to the drain lead ($I_Q^r$). This ratio between the two heat currents is known as the heat current amplification factor ($\alpha$) and serves as the hallmark of the thermal transistor effect. (c) The heat current amplification factor $\alpha$ as a function of QD energy difference $E_l-E_r$ for the cases of coherence and incoherence. The other parameters are $\mu=6\, {\rm meV}$, $\Delta\mu = 0$,  $E_l=12\, {\rm meV}$, $\Gamma_l=\Gamma_r=\Gamma_{\rm ph}=6 \, {\rm meV}$, $t=8 \, {\rm meV}$, $k_BT_l=k_BT_r=k_BT_{\rm ph}=10\, {\rm meV}$. (d) The heat current amplification factor $\alpha$ as functions of QD energy $E_l$ and $E_r$ The other parameters are same with Fig.~\ref{Fig:transistor}(c). }
\label{Fig:transistor}
\end{center}
\end{figure}

{\emph{Thermal transistor effect in the linear-response regime.}} Heat amplification, a phenomenon where a minor adjustment in the base current yields a substantial impact on the current flowing between the collector and emitter, has led to the development of quantum thermal transistors, enabling precise control over energy transport~\cite{Transistor9,YuPRE18,ZhangAPL23}. As depicted in Fig.\ref{Fig:transistor}(a), this schematic outlines the operation of a conventional transistor. In our research, we investigate three-terminal quantum-dot systems as thermal transistors, as depicted in Fig.\ref{Fig:transistor}(b).

Traditionally, it was widely accepted that nonlinear transport was a prerequisite for thermal transistor effects. Negative differential thermal conductance was considered a key condition for the emergence of such effects, as outlined in~\cite{RenRMP,LiPRL04,LiAPL06}. However, a groundbreaking perspective was introduced in Ref.\cite{Jiangtransistors}, proposing that thermal transistor effects could occur in the linear-transport regime, particularly when dominated by phonon-assisted inelastic transport. In our subsequent work~\cite{MyPRBdiode,LuEntropy}, we further demonstrated the existence of linear thermal transistor effects in circuit-quantum-electrodynamics systems across a wide range of QD energies and light-matter interactions. Furthermore, we showed that thermal transistor effects can manifest even without these conditions in Brownian thermal transistors. Analyzing the statistical distributions of the thermal amplification factor within the Gaussian fluctuation framework provided further insights~\cite{MyPRBtransistor}. However, it's important to note that all these studies were limited to the quantum incoherent regime, and the influence of quantum coherence on the thermal transistor effect remains unexplored.

In the scenario of exclusively thermal conduction (with the electrochemical potential difference set to zero), the system's linear thermal transport properties can be aptly described by~\cite{Transistor1,WangPRA18}   
\begin{equation}
\begin{aligned}
\left( \begin{array}{cccc} I_Q^r\\ I_Q^{\rm ph} \end{array}\right) =
 \left( \begin{array}{cccc} K_{RR} & K_{RP} \\ K_{PR} &
    K_{PP} \end{array} \right) \left( \begin{array}{cccc} T_r-T_l\\
    T_{\rm ph}-T_l \end{array}\right),
\end{aligned}
\end{equation}
One of our key findings is the potential for a thermal transistor effect to emerge within the linear-response regime. To be more specific, the amplification factor for heat current is defined as follows:  
\begin{equation}
\begin{aligned}
\alpha=\Big|\frac{\partial_{T_{\rm ph}}I_Q^r}{\partial_{T_{\rm ph}}I_Q^P}\Big|=\frac{K_{RP}}{K_{PP}},
\end{aligned}
\end{equation}

As demonstrated in Figures~\ref{Fig:transistor}(c) and \ref{Fig:transistor}(d), we have observed that the coefficient $\alpha$ can be greater than 1 due to the inelastic transport process and displays high sensitivity to the quantum dot energies, $E_l$ and $E_r$, which can be easily manipulated through gate voltages in experimental setups. Initially, we observed that the thermal amplification effect could be achieved within a specific parameter range when 
$(E_r - E_l) > 8  \mathrm{meV}$ in the context of incoherent transport. However, when accounting for quantum coherence, while thermal amplification effects remain achievable, the range where $\alpha > 1$ becomes notably restricted, specifically within $11  \mathrm{meV} < (E_r - E_l) < 13  \mathrm{meV}$. According to the findings in Ref.\cite{Jiangtransistors}, the thermal amplification coefficient $\alpha$ is determined as $\alpha = |E_l / (E_r - E_l)|$ in the regime of weak electron-phonon coupling using the Fermi Golden rule method. The primary driver of the transistor effect is the ratio of incoherent thermal conductivity. Quantum coherence introduces interference effects, increases sensitivity to energy levels, and is susceptible to decoherence and noise, all of which narrow the parameter space for achieving the thermal transistor effect. 
Consequently, though quantum coherence significantly impacts currents as illustrated in Fig.~\ref{fig:3Tsystem}(b), it does not facilitate the thermal transistor effect in the linear-transport regime, limiting rather than enhancing the conditions under which thermal amplification is possible. In summary, quantum coherence effects, while crucial for understanding the dynamics of quantum dot systems, restrict the thermal transistor effect in the linear response regime.

\emph{Conclusion.} In conclusion, this study delves into the intriguing domain of inelastic thermoelectric devices, revealing their potential as quantum thermoelectric diodes and transistors via inelastic transport processes induced by electron-phonon interactions. We have demonstrated that precise adjustment of QD energy levels can yield significant charge and Seebeck rectification effects, which are further enhanced when quantum coherence is considered. Additionally, our research has unveiled the capability of a three-terminal configuration to produce thermal transistor effects due to the inelastic transport, even within the linear transport regime. However, it's important to note that quantum coherence does not contribute to heat amplification effect. In summary, our study underscores the significance of quantum coherence in the realm of inelastic thermoelectric devices and opens the door to innovative applications in the field of quantum thermoelectricity.

{\emph{Acknowledgments.}} We thank for Professor Jian-Hua Jiang, Professor Rafael S\'anchez for helpful discussions. This work was supported by the funding for the National Natural Science Foundation of China under Grant No. 12305050, Jiangsu Key Disciplines of the Fourteenth Five-Year Plan (Grant No. 2021135), the Natural Science Foundation of Jiangsu Higher Education Institutions of China (Grant No. 23KJB140017), and the Opening Project of Shanghai Key Laboratory of Special Artificial Microstructure Materials and Technology. 

\bibliography{Ref-diode}

\begin{thebibliography}{55}%
\makeatletter
\providecommand \@ifxundefined [1]{%
 \@ifx{#1\undefined}
}%
\providecommand \@ifnum [1]{%
 \ifnum #1\expandafter \@firstoftwo
 \else \expandafter \@secondoftwo
 \fi
}%
\providecommand \@ifx [1]{%
 \ifx #1\expandafter \@firstoftwo
 \else \expandafter \@secondoftwo
 \fi
}%
\providecommand \natexlab [1]{#1}%
\providecommand \enquote  [1]{``#1''}%
\providecommand \bibnamefont  [1]{#1}%
\providecommand \bibfnamefont [1]{#1}%
\providecommand \citenamefont [1]{#1}%
\providecommand \href@noop [0]{\@secondoftwo}%
\providecommand \href [0]{\begingroup \@sanitize@url \@href}%
\providecommand \@href[1]{\@@startlink{#1}\@@href}%
\providecommand \@@href[1]{\endgroup#1\@@endlink}%
\providecommand \@sanitize@url [0]{\catcode `\\12\catcode `\$12\catcode
  `\&12\catcode `\#12\catcode `\^12\catcode `\_12\catcode `\%12\relax}%
\providecommand \@@startlink[1]{}%
\providecommand \@@endlink[0]{}%
\providecommand \url  [0]{\begingroup\@sanitize@url \@url }%
\providecommand \@url [1]{\endgroup\@href {#1}{\urlprefix }}%
\providecommand \urlprefix  [0]{URL }%
\providecommand \Eprint [0]{\href }%
\providecommand \doibase [0]{http://dx.doi.org/}%
\providecommand \selectlanguage [0]{\@gobble}%
\providecommand \bibinfo  [0]{\@secondoftwo}%
\providecommand \bibfield  [0]{\@secondoftwo}%
\providecommand \translation [1]{[#1]}%
\providecommand \BibitemOpen [0]{}%
\providecommand \bibitemStop [0]{}%
\providecommand \bibitemNoStop [0]{.\EOS\space}%
\providecommand \EOS [0]{\spacefactor3000\relax}%
\providecommand \BibitemShut  [1]{\csname bibitem#1\endcsname}%
\let\auto@bib@innerbib\@empty
\bibitem [{\citenamefont {Li}\ \emph {et~al.}(2012)\citenamefont {Li},
  \citenamefont {Ren}, \citenamefont {Wang}, \citenamefont {Zhang},
  \citenamefont {H\"anggi},\ and\ \citenamefont {Li}}]{RenRMP}%
  \BibitemOpen
  \bibfield  {author} {\bibinfo {author} {\bibfnamefont {N.}~\bibnamefont
  {Li}}, \bibinfo {author} {\bibfnamefont {J.}~\bibnamefont {Ren}}, \bibinfo
  {author} {\bibfnamefont {L.}~\bibnamefont {Wang}}, \bibinfo {author}
  {\bibfnamefont {G.}~\bibnamefont {Zhang}}, \bibinfo {author} {\bibfnamefont
  {P.}~\bibnamefont {H\"anggi}}, \ and\ \bibinfo {author} {\bibfnamefont
  {B.}~\bibnamefont {Li}},\ }\bibfield  {title} {\enquote {\bibinfo {title}
  {Phononics: Manipulating heat flow with electronic analogs and beyond},}\
  }\href {\doibase 10.1103/RevModPhys.84.1045} {\bibfield  {journal} {\bibinfo
  {journal} {Rev. Mod. Phys.}\ }\textbf {\bibinfo {volume} {84}},\ \bibinfo
  {pages} {1045--1066} (\bibinfo {year} {2012})}\BibitemShut {NoStop}%
\bibitem [{\citenamefont {Thierschmann}\ \emph {et~al.}(2015)\citenamefont
  {Thierschmann}, \citenamefont {S{\'a}nchez}, \citenamefont {Sothmann},
  \citenamefont {Arnold}, \citenamefont {Heyn}, \citenamefont {Hansen},
  \citenamefont {Buhmann},\ and\ \citenamefont {Molenkamp}}]{Thier2015}%
  \BibitemOpen
  \bibfield  {author} {\bibinfo {author} {\bibfnamefont {H.}~\bibnamefont
  {Thierschmann}}, \bibinfo {author} {\bibfnamefont {R.}~\bibnamefont
  {S{\'a}nchez}}, \bibinfo {author} {\bibfnamefont {B.}~\bibnamefont
  {Sothmann}}, \bibinfo {author} {\bibfnamefont {F.}~\bibnamefont {Arnold}},
  \bibinfo {author} {\bibfnamefont {C.}~\bibnamefont {Heyn}}, \bibinfo {author}
  {\bibfnamefont {W.}~\bibnamefont {Hansen}}, \bibinfo {author} {\bibfnamefont
  {H.}~\bibnamefont {Buhmann}}, \ and\ \bibinfo {author} {\bibfnamefont
  {L.~W.}\ \bibnamefont {Molenkamp}},\ }\bibfield  {title} {\enquote {\bibinfo
  {title} {Three-terminal energy harvester with coupled quantum dots},}\ }\href
  {\doibase 10.1038/nnano.2015.176} {\bibfield  {journal} {\bibinfo  {journal}
  {Nat. Nanotech.}\ }\textbf {\bibinfo {volume} {10}},\ \bibinfo {pages} {854}
  (\bibinfo {year} {2015})}\BibitemShut {NoStop}%
\bibitem [{\citenamefont {Benenti}\ \emph {et~al.}(2017)\citenamefont
  {Benenti}, \citenamefont {Casati}, \citenamefont {Saito},\ and\ \citenamefont
  {Whitney}}]{BenentiPR}%
  \BibitemOpen
  \bibfield  {author} {\bibinfo {author} {\bibfnamefont {G.}~\bibnamefont
  {Benenti}}, \bibinfo {author} {\bibfnamefont {G.}~\bibnamefont {Casati}},
  \bibinfo {author} {\bibfnamefont {K.}~\bibnamefont {Saito}}, \ and\ \bibinfo
  {author} {\bibfnamefont {R.~S.}\ \bibnamefont {Whitney}},\ }\bibfield
  {title} {\enquote {\bibinfo {title} {Fundamental aspects of steady-state
  conversion of heat to work at the nanoscale},}\ }\href {\doibase
  10.1016/j.physrep.2017.05.008} {\bibfield  {journal} {\bibinfo  {journal}
  {Phys. Rep.}\ }\textbf {\bibinfo {volume} {694}},\ \bibinfo {pages} {1}
  (\bibinfo {year} {2017})}\BibitemShut {NoStop}%
\bibitem [{\citenamefont {Jiang}\ \emph {et~al.}(2012)\citenamefont {Jiang},
  \citenamefont {Entin-Wohlman},\ and\ \citenamefont {Imry}}]{Jiang2012}%
  \BibitemOpen
  \bibfield  {author} {\bibinfo {author} {\bibfnamefont {J.-H.}\ \bibnamefont
  {Jiang}}, \bibinfo {author} {\bibfnamefont {O.}~\bibnamefont
  {Entin-Wohlman}}, \ and\ \bibinfo {author} {\bibfnamefont {Y.}~\bibnamefont
  {Imry}},\ }\bibfield  {title} {\enquote {\bibinfo {title} {Thermoelectric
  three-terminal hopping transport through one-dimensional nanosystems},}\
  }\href {\doibase 10.1103/PhysRevB.85.075412} {\bibfield  {journal} {\bibinfo
  {journal} {Phys. Rev. B}\ }\textbf {\bibinfo {volume} {85}},\ \bibinfo
  {pages} {075412} (\bibinfo {year} {2012})}\BibitemShut {NoStop}%
\bibitem [{\citenamefont {Jiang}\ and\ \citenamefont {Imry}(2017)}]{Jiang2017}%
  \BibitemOpen
  \bibfield  {author} {\bibinfo {author} {\bibfnamefont {J.-H.}\ \bibnamefont
  {Jiang}}\ and\ \bibinfo {author} {\bibfnamefont {Y.}~\bibnamefont {Imry}},\
  }\bibfield  {title} {\enquote {\bibinfo {title} {Enhancing thermoelectric
  performance using nonlinear transport effects},}\ }\href {\doibase
  10.1103/PhysRevApplied.7.064001} {\bibfield  {journal} {\bibinfo  {journal}
  {Phys. Rev. Applied}\ }\textbf {\bibinfo {volume} {7}},\ \bibinfo {pages}
  {064001} (\bibinfo {year} {2017})}\BibitemShut {NoStop}%
\bibitem [{\citenamefont {Liu}\ \emph {et~al.}(2019)\citenamefont {Liu},
  \citenamefont {Wang}, \citenamefont {Wang},\ and\ \citenamefont
  {Ren}}]{WangPRE}%
  \BibitemOpen
  \bibfield  {author} {\bibinfo {author} {\bibfnamefont {H.}~\bibnamefont
  {Liu}}, \bibinfo {author} {\bibfnamefont {C.}~\bibnamefont {Wang}}, \bibinfo
  {author} {\bibfnamefont {L.-Q.}\ \bibnamefont {Wang}}, \ and\ \bibinfo
  {author} {\bibfnamefont {J.}~\bibnamefont {Ren}},\ }\bibfield  {title}
  {\enquote {\bibinfo {title} {Strong system-bath coupling induces negative
  differential thermal conductance and heat amplification in nonequilibrium
  two-qubit systems},}\ }\href {\doibase 10.1103/PhysRevE.99.032114} {\bibfield
   {journal} {\bibinfo  {journal} {Phys. Rev. E}\ }\textbf {\bibinfo {volume}
  {99}},\ \bibinfo {pages} {032114} (\bibinfo {year} {2019})}\BibitemShut
  {NoStop}%
\bibitem [{\citenamefont {Lu}\ \emph {et~al.}(2023)\citenamefont {Lu},
  \citenamefont {Wang}, \citenamefont {Wang},\ and\ \citenamefont
  {Jiang}}]{LuEntropy}%
  \BibitemOpen
  \bibfield  {author} {\bibinfo {author} {\bibfnamefont {J.}~\bibnamefont
  {Lu}}, \bibinfo {author} {\bibfnamefont {R.}~\bibnamefont {Wang}}, \bibinfo
  {author} {\bibfnamefont {C.}~\bibnamefont {Wang}}, \ and\ \bibinfo {author}
  {\bibfnamefont {J.-H.}\ \bibnamefont {Jiang}},\ }\bibfield  {title} {\enquote
  {\bibinfo {title} {Thermoelectric rectification and amplification in
  interacting quantum-dot circuit-quantum-electrodynamics systems},}\ }\href
  {\doibase 10.3390/e25030498} {\bibfield  {journal} {\bibinfo  {journal}
  {Entropy}\ }\textbf {\bibinfo {volume} {25}} (\bibinfo {year} {2023}),\
  10.3390/e25030498}\BibitemShut {NoStop}%
\bibitem [{\citenamefont {Ren}(2023)}]{ChinPhysLett.40.090501}%
  \BibitemOpen
  \bibfield  {author} {\bibinfo {author} {\bibfnamefont {Jie}\ \bibnamefont
  {Ren}},\ }\bibfield  {title} {\enquote {\bibinfo {title} {Geometric
  thermoelectric pump: Energy harvesting beyond seebeck and pyroelectric
  effects},}\ }\href {\doibase 10.1088/0256-307X/40/9/090501} {\bibfield
  {journal} {\bibinfo  {journal} {Chin. Phys. Lett.}\ }\textbf {\bibinfo
  {volume} {40}},\ \bibinfo {pages} {090501} (\bibinfo {year}
  {2023})}\BibitemShut {NoStop}%
\bibitem [{\citenamefont {Bu}\ \emph {et~al.}(2017)\citenamefont {Bu},
  \citenamefont {Singh}, \citenamefont {Fei}, \citenamefont {Pati},\ and\
  \citenamefont {Wu}}]{WuPRL17}%
  \BibitemOpen
  \bibfield  {author} {\bibinfo {author} {\bibfnamefont {K.}~\bibnamefont
  {Bu}}, \bibinfo {author} {\bibfnamefont {U.}~\bibnamefont {Singh}}, \bibinfo
  {author} {\bibfnamefont {S.-M.}\ \bibnamefont {Fei}}, \bibinfo {author}
  {\bibfnamefont {A.~K.}\ \bibnamefont {Pati}}, \ and\ \bibinfo {author}
  {\bibfnamefont {J.}~\bibnamefont {Wu}},\ }\bibfield  {title} {\enquote
  {\bibinfo {title} {Maximum relative entropy of coherence: An operational
  coherence measure},}\ }\href {\doibase 10.1103/PhysRevLett.119.150405}
  {\bibfield  {journal} {\bibinfo  {journal} {Phys. Rev. Lett.}\ }\textbf
  {\bibinfo {volume} {119}},\ \bibinfo {pages} {150405} (\bibinfo {year}
  {2017})}\BibitemShut {NoStop}%
\bibitem [{\citenamefont {Camati}\ \emph {et~al.}(2019)\citenamefont {Camati},
  \citenamefont {Santos},\ and\ \citenamefont {Serra}}]{CamatiPRA19}%
  \BibitemOpen
  \bibfield  {author} {\bibinfo {author} {\bibfnamefont {P.~A.}\ \bibnamefont
  {Camati}}, \bibinfo {author} {\bibfnamefont {J.~F.~G.}\ \bibnamefont
  {Santos}}, \ and\ \bibinfo {author} {\bibfnamefont {R.~M.}\ \bibnamefont
  {Serra}},\ }\bibfield  {title} {\enquote {\bibinfo {title} {Coherence effects
  in the performance of the quantum otto heat engine},}\ }\href {\doibase
  10.1103/PhysRevA.99.062103} {\bibfield  {journal} {\bibinfo  {journal} {Phys.
  Rev. A}\ }\textbf {\bibinfo {volume} {99}},\ \bibinfo {pages} {062103}
  (\bibinfo {year} {2019})}\BibitemShut {NoStop}%
\bibitem [{\citenamefont {Liu}\ and\ \citenamefont {Segal}(2021)}]{SegalPRE21}%
  \BibitemOpen
  \bibfield  {author} {\bibinfo {author} {\bibfnamefont {J.}~\bibnamefont
  {Liu}}\ and\ \bibinfo {author} {\bibfnamefont {D.}~\bibnamefont {Segal}},\
  }\bibfield  {title} {\enquote {\bibinfo {title} {Coherences and the
  thermodynamic uncertainty relation: Insights from quantum absorption
  refrigerators},}\ }\href {\doibase 10.1103/PhysRevE.103.032138} {\bibfield
  {journal} {\bibinfo  {journal} {Phys. Rev. E}\ }\textbf {\bibinfo {volume}
  {103}},\ \bibinfo {pages} {032138} (\bibinfo {year} {2021})}\BibitemShut
  {NoStop}%
\bibitem [{\citenamefont {Aimet}\ and\ \citenamefont {Kwon}(2023)}]{KwonPRA23}%
  \BibitemOpen
  \bibfield  {author} {\bibinfo {author} {\bibfnamefont {S.}~\bibnamefont
  {Aimet}}\ and\ \bibinfo {author} {\bibfnamefont {H.}~\bibnamefont {Kwon}},\
  }\bibfield  {title} {\enquote {\bibinfo {title} {Engineering a heat engine
  purely driven by quantum coherence},}\ }\href {\doibase
  10.1103/PhysRevA.107.012221} {\bibfield  {journal} {\bibinfo  {journal}
  {Phys. Rev. A}\ }\textbf {\bibinfo {volume} {107}},\ \bibinfo {pages}
  {012221} (\bibinfo {year} {2023})}\BibitemShut {NoStop}%
\bibitem [{\citenamefont {Behera}\ \emph {et~al.}(2023)\citenamefont {Behera},
  \citenamefont {Bedkihal}, \citenamefont {Agarwalla},\ and\ \citenamefont
  {Bandyopadhyay}}]{BeheraPRB23}%
  \BibitemOpen
  \bibfield  {author} {\bibinfo {author} {\bibfnamefont {J.}~\bibnamefont
  {Behera}}, \bibinfo {author} {\bibfnamefont {S.}~\bibnamefont {Bedkihal}},
  \bibinfo {author} {\bibfnamefont {B.~K.}\ \bibnamefont {Agarwalla}}, \ and\
  \bibinfo {author} {\bibfnamefont {M.}~\bibnamefont {Bandyopadhyay}},\
  }\bibfield  {title} {\enquote {\bibinfo {title} {Quantum coherent control of
  nonlinear thermoelectric transport in a triple-dot aharonov-bohm heat
  engine},}\ }\href {\doibase 10.1103/PhysRevB.108.165419} {\bibfield
  {journal} {\bibinfo  {journal} {Phys. Rev. B}\ }\textbf {\bibinfo {volume}
  {108}},\ \bibinfo {pages} {165419} (\bibinfo {year} {2023})}\BibitemShut
  {NoStop}%
\bibitem [{\citenamefont {Chen}\ \emph {et~al.}(2023)\citenamefont {Chen},
  \citenamefont {Wang}, \citenamefont {Chen}, \citenamefont {Lu},\ and\
  \citenamefont {Wang}}]{ChinPhysLett.40.050501}%
  \BibitemOpen
  \bibfield  {author} {\bibinfo {author} {\bibfnamefont {Ze-Huan}\ \bibnamefont
  {Chen}}, \bibinfo {author} {\bibfnamefont {Fei-Yu}\ \bibnamefont {Wang}},
  \bibinfo {author} {\bibfnamefont {Hua}\ \bibnamefont {Chen}}, \bibinfo
  {author} {\bibfnamefont {Jin-Cheng}\ \bibnamefont {Lu}}, \ and\ \bibinfo
  {author} {\bibfnamefont {Chen}\ \bibnamefont {Wang}},\ }\bibfield  {title}
  {\enquote {\bibinfo {title} {Modulation of steady-state heat transport in a
  dissipative multi-mode qubit-photon system},}\ }\href {\doibase
  10.1088/0256-307X/40/5/050501} {\bibfield  {journal} {\bibinfo  {journal}
  {Chin. Phys. Lett.}\ }\textbf {\bibinfo {volume} {40}},\ \bibinfo {pages}
  {050501} (\bibinfo {year} {2023})}\BibitemShut {NoStop}%
\bibitem [{\citenamefont {Zhang}\ \emph {et~al.}(2023)\citenamefont {Zhang},
  \citenamefont {Li}, \citenamefont {Xu},\ and\ \citenamefont
  {Huang}}]{ChinPhysLett.40.054401}%
  \BibitemOpen
  \bibfield  {author} {\bibinfo {author} {\bibfnamefont {Chuan-Xin}\
  \bibnamefont {Zhang}}, \bibinfo {author} {\bibfnamefont {Tian-Jiao}\
  \bibnamefont {Li}}, \bibinfo {author} {\bibfnamefont {Liu-Jun}\ \bibnamefont
  {Xu}}, \ and\ \bibinfo {author} {\bibfnamefont {Ji-Ping}\ \bibnamefont
  {Huang}},\ }\bibfield  {title} {\enquote {\bibinfo {title} {Dust-induced
  regulation of thermal radiation in water droplets},}\ }\href {\doibase
  10.1088/0256-307X/40/5/054401} {\bibfield  {journal} {\bibinfo  {journal}
  {Chin. Phys. Lett.}\ }\textbf {\bibinfo {volume} {40}},\ \bibinfo {pages}
  {054401} (\bibinfo {year} {2023})}\BibitemShut {NoStop}%
\bibitem [{\citenamefont {Engel}\ \emph {et~al.}(2007)\citenamefont {Engel},
  \citenamefont {Calhoun}, \citenamefont {Read}, \citenamefont {Ahn},
  \citenamefont {Man{\v{c}}al}, \citenamefont {Cheng}, \citenamefont
  {Blankenship},\ and\ \citenamefont {Fleming}}]{EngelNature}%
  \BibitemOpen
  \bibfield  {author} {\bibinfo {author} {\bibfnamefont {G.~S.}\ \bibnamefont
  {Engel}}, \bibinfo {author} {\bibfnamefont {T.~R.}\ \bibnamefont {Calhoun}},
  \bibinfo {author} {\bibfnamefont {E.~L.}\ \bibnamefont {Read}}, \bibinfo
  {author} {\bibfnamefont {T.-K.}\ \bibnamefont {Ahn}}, \bibinfo {author}
  {\bibfnamefont {T.}~\bibnamefont {Man{\v{c}}al}}, \bibinfo {author}
  {\bibfnamefont {Y.-C.}\ \bibnamefont {Cheng}}, \bibinfo {author}
  {\bibfnamefont {R.~E.}\ \bibnamefont {Blankenship}}, \ and\ \bibinfo {author}
  {\bibfnamefont {G.~R.}\ \bibnamefont {Fleming}},\ }\bibfield  {title}
  {\enquote {\bibinfo {title} {Evidence for wavelike energy transfer through
  quantum coherence in photosynthetic systems},}\ }\href {\doibase
  10.1038/nature05678} {\bibfield  {journal} {\bibinfo  {journal} {Nature}\
  }\textbf {\bibinfo {volume} {446}},\ \bibinfo {pages} {782} (\bibinfo {year}
  {2007})}\BibitemShut {NoStop}%
\bibitem [{\citenamefont {V.}\ and\ \citenamefont {P.}(2010)}]{Vaziri10}%
  \BibitemOpen
  \bibfield  {author} {\bibinfo {author} {\bibfnamefont {Alipasha}\
  \bibnamefont {V.}}\ and\ \bibinfo {author} {\bibfnamefont {Martin~B.}\
  \bibnamefont {P.}},\ }\bibfield  {title} {\enquote {\bibinfo {title} {Quantum
  coherence in ion channels: resonances, transport and verification},}\ }\href
  {\doibase 10.1088/1367-2630/12/8/085001} {\bibfield  {journal} {\bibinfo
  {journal} {New J. Phys.}\ }\textbf {\bibinfo {volume} {12}},\ \bibinfo
  {pages} {085001} (\bibinfo {year} {2010})}\BibitemShut {NoStop}%
\bibitem [{\citenamefont {S\'anchez}\ \emph {et~al.}(2021)\citenamefont
  {S\'anchez}, \citenamefont {Gorini},\ and\ \citenamefont
  {Fleury}}]{SanchezPRB22}%
  \BibitemOpen
  \bibfield  {author} {\bibinfo {author} {\bibfnamefont {R.}~\bibnamefont
  {S\'anchez}}, \bibinfo {author} {\bibfnamefont {C.}~\bibnamefont {Gorini}}, \
  and\ \bibinfo {author} {\bibfnamefont {G.}~\bibnamefont {Fleury}},\
  }\bibfield  {title} {\enquote {\bibinfo {title} {Extrinsic thermoelectric
  response of coherent conductors},}\ }\href {\doibase
  10.1103/PhysRevB.104.115430} {\bibfield  {journal} {\bibinfo  {journal}
  {Phys. Rev. B}\ }\textbf {\bibinfo {volume} {104}},\ \bibinfo {pages}
  {115430} (\bibinfo {year} {2021})}\BibitemShut {NoStop}%
\bibitem [{\citenamefont {Ivander}\ \emph {et~al.}(2022)\citenamefont
  {Ivander}, \citenamefont {Anto-Sztrikacs},\ and\ \citenamefont
  {Segal}}]{IvanderNJP22}%
  \BibitemOpen
  \bibfield  {author} {\bibinfo {author} {\bibfnamefont {F.}~\bibnamefont
  {Ivander}}, \bibinfo {author} {\bibfnamefont {N.}~\bibnamefont
  {Anto-Sztrikacs}}, \ and\ \bibinfo {author} {\bibfnamefont {D.}~\bibnamefont
  {Segal}},\ }\bibfield  {title} {\enquote {\bibinfo {title} {Quantum
  coherence-control of thermal energy transport: the v model as a case
  study},}\ }\href {\doibase 10.1088/1367-2630/ac9498} {\bibfield  {journal}
  {\bibinfo  {journal} {New J. Phys.}\ }\textbf {\bibinfo {volume} {24}},\
  \bibinfo {pages} {103010} (\bibinfo {year} {2022})}\BibitemShut {NoStop}%
\bibitem [{\citenamefont {Yuan}\ \emph {et~al.}(2023)\citenamefont {Yuan},
  \citenamefont {Ruan}, \citenamefont {Liu}, \citenamefont {He},\ and\
  \citenamefont {Wang}}]{Yuan2023}%
  \BibitemOpen
  \bibfield  {author} {\bibinfo {author} {\bibfnamefont {Jiehong}\ \bibnamefont
  {Yuan}}, \bibinfo {author} {\bibfnamefont {Huilin}\ \bibnamefont {Ruan}},
  \bibinfo {author} {\bibfnamefont {Dehua}\ \bibnamefont {Liu}}, \bibinfo
  {author} {\bibfnamefont {Jizhou}\ \bibnamefont {He}}, \ and\ \bibinfo
  {author} {\bibfnamefont {Jianhui}\ \bibnamefont {Wang}},\ }\bibfield  {title}
  {\enquote {\bibinfo {title} {Quantum brayton refrigeration cycle with
  finite-size bose–einstein condensates},}\ }\href {\doibase
  10.1088/0256-307X/40/10/100502} {\bibfield  {journal} {\bibinfo  {journal}
  {Chinese Physics Letters}\ }\textbf {\bibinfo {volume} {40}},\ \bibinfo
  {pages} {100502} (\bibinfo {year} {2023})}\BibitemShut {NoStop}%
\bibitem [{\citenamefont {Gao}\ \emph {et~al.}(2023)\citenamefont {Gao},
  \citenamefont {Liu}, \citenamefont {Wang},\ and\ \citenamefont
  {He}}]{ChinPhysLett.40.117301}%
  \BibitemOpen
  \bibfield  {author} {\bibinfo {author} {\bibfnamefont {Jin-Zhu}\ \bibnamefont
  {Gao}}, \bibinfo {author} {\bibfnamefont {Xing}\ \bibnamefont {Liu}},
  \bibinfo {author} {\bibfnamefont {Jian-Hui}\ \bibnamefont {Wang}}, \ and\
  \bibinfo {author} {\bibfnamefont {Ji-Zhou}\ \bibnamefont {He}},\ }\bibfield
  {title} {\enquote {\bibinfo {title} {Cooling by coulomb heat drag based on
  three coupled quantum dots},}\ }\href {\doibase
  10.1088/0256-307X/40/11/117301} {\bibfield  {journal} {\bibinfo  {journal}
  {Chin. Phys. Lett.}\ }\textbf {\bibinfo {volume} {40}},\ \bibinfo {pages}
  {117301} (\bibinfo {year} {2023})}\BibitemShut {NoStop}%
\bibitem [{\citenamefont {Wang}\ \emph
  {et~al.}(2022{\natexlab{a}})\citenamefont {Wang}, \citenamefont {Wang},
  \citenamefont {Lu},\ and\ \citenamefont {Jiang}}]{MyReview}%
  \BibitemOpen
  \bibfield  {author} {\bibinfo {author} {\bibfnamefont {R.}~\bibnamefont
  {Wang}}, \bibinfo {author} {\bibfnamefont {C.}~\bibnamefont {Wang}}, \bibinfo
  {author} {\bibfnamefont {J.}~\bibnamefont {Lu}}, \ and\ \bibinfo {author}
  {\bibfnamefont {J.-H.}\ \bibnamefont {Jiang}},\ }\bibfield  {title} {\enquote
  {\bibinfo {title} {Inelastic thermoelectric transport and fluctuations in
  mesoscopic systems},}\ }\href {\doibase 10.1080/23746149.2022.2082317}
  {\bibfield  {journal} {\bibinfo  {journal} {Adv. Phys.: X}\ }\textbf
  {\bibinfo {volume} {7}},\ \bibinfo {pages} {2082317} (\bibinfo {year}
  {2022}{\natexlab{a}})}\BibitemShut {NoStop}%
\bibitem [{\citenamefont {Entin-Wohlman}\ \emph {et~al.}(2010)\citenamefont
  {Entin-Wohlman}, \citenamefont {Imry},\ and\ \citenamefont
  {Aharony}}]{OraPRB2010}%
  \BibitemOpen
  \bibfield  {author} {\bibinfo {author} {\bibfnamefont {O.}~\bibnamefont
  {Entin-Wohlman}}, \bibinfo {author} {\bibfnamefont {Y.}~\bibnamefont {Imry}},
  \ and\ \bibinfo {author} {\bibfnamefont {A.}~\bibnamefont {Aharony}},\
  }\bibfield  {title} {\enquote {\bibinfo {title} {Three-terminal
  thermoelectric transport through a molecular junction},}\ }\href {\doibase
  10.1103/PhysRevB.82.115314} {\bibfield  {journal} {\bibinfo  {journal} {Phys.
  Rev. B}\ }\textbf {\bibinfo {volume} {82}},\ \bibinfo {pages} {115314}
  (\bibinfo {year} {2010})}\BibitemShut {NoStop}%
\bibitem [{\citenamefont {Lu}\ \emph {et~al.}(2021)\citenamefont {Lu},
  \citenamefont {Jiang},\ and\ \citenamefont {Imry}}]{MyPRBdemon}%
  \BibitemOpen
  \bibfield  {author} {\bibinfo {author} {\bibfnamefont {J.}~\bibnamefont
  {Lu}}, \bibinfo {author} {\bibfnamefont {J.-H.}\ \bibnamefont {Jiang}}, \
  and\ \bibinfo {author} {\bibfnamefont {Y.}~\bibnamefont {Imry}},\ }\bibfield
  {title} {\enquote {\bibinfo {title} {Unconventional four-terminal
  thermoelectric transport due to inelastic transport: Cooling by transverse
  heat current, transverse thermoelectric effect, and maxwell demon},}\ }\href
  {\doibase 10.1103/PhysRevB.103.085429} {\bibfield  {journal} {\bibinfo
  {journal} {Phys. Rev. B}\ }\textbf {\bibinfo {volume} {103}},\ \bibinfo
  {pages} {085429} (\bibinfo {year} {2021})}\BibitemShut {NoStop}%
\bibitem [{\citenamefont {Xi}\ \emph {et~al.}(2021)\citenamefont {Xi},
  \citenamefont {Wang}, \citenamefont {Lu},\ and\ \citenamefont
  {Jiang}}]{Xi21CPL}%
  \BibitemOpen
  \bibfield  {author} {\bibinfo {author} {\bibfnamefont {M.}~\bibnamefont
  {Xi}}, \bibinfo {author} {\bibfnamefont {R.}~\bibnamefont {Wang}}, \bibinfo
  {author} {\bibfnamefont {J.}~\bibnamefont {Lu}}, \ and\ \bibinfo {author}
  {\bibfnamefont {J.-H.}\ \bibnamefont {Jiang}},\ }\bibfield  {title} {\enquote
  {\bibinfo {title} {Coulomb thermoelectric drag in four-terminal mesoscopic
  quantum transport},}\ }\href {\doibase 10.1088/0256-307X/38/8/088801}
  {\bibfield  {journal} {\bibinfo  {journal} {Chin. Phys. Lett.}\ }\textbf
  {\bibinfo {volume} {38}},\ \bibinfo {pages} {088801} (\bibinfo {year}
  {2021})}\BibitemShut {NoStop}%
\bibitem [{\citenamefont {Nian}\ \emph
  {et~al.}(2023{\natexlab{a}})\citenamefont {Nian}, \citenamefont {Hu},
  \citenamefont {Xiong}, \citenamefont {L\"u},\ and\ \citenamefont
  {Zheng}}]{PhysRevB.108.085430}%
  \BibitemOpen
  \bibfield  {author} {\bibinfo {author} {\bibfnamefont {Lei-Lei}\ \bibnamefont
  {Nian}}, \bibinfo {author} {\bibfnamefont {Shiqian}\ \bibnamefont {Hu}},
  \bibinfo {author} {\bibfnamefont {Long}\ \bibnamefont {Xiong}}, \bibinfo
  {author} {\bibfnamefont {Jing-Tao}\ \bibnamefont {L\"u}}, \ and\ \bibinfo
  {author} {\bibfnamefont {Bo}~\bibnamefont {Zheng}},\ }\bibfield  {title}
  {\enquote {\bibinfo {title} {Photon-assisted electron transport across a
  quantum phase transition},}\ }\href {\doibase 10.1103/PhysRevB.108.085430}
  {\bibfield  {journal} {\bibinfo  {journal} {Phys. Rev. B}\ }\textbf {\bibinfo
  {volume} {108}},\ \bibinfo {pages} {085430} (\bibinfo {year}
  {2023}{\natexlab{a}})}\BibitemShut {NoStop}%
\bibitem [{\citenamefont {Nian}\ \emph
  {et~al.}(2023{\natexlab{b}})\citenamefont {Nian}, \citenamefont {Zheng},\
  and\ \citenamefont {L\"u}}]{PhysRevB.107.L241405}%
  \BibitemOpen
  \bibfield  {author} {\bibinfo {author} {\bibfnamefont {Lei-Lei}\ \bibnamefont
  {Nian}}, \bibinfo {author} {\bibfnamefont {Bo}~\bibnamefont {Zheng}}, \ and\
  \bibinfo {author} {\bibfnamefont {Jing-Tao}\ \bibnamefont {L\"u}},\
  }\bibfield  {title} {\enquote {\bibinfo {title} {Electrically driven photon
  statistics engineering in quantum-dot circuit quantum electrodynamics},}\
  }\href {\doibase 10.1103/PhysRevB.107.L241405} {\bibfield  {journal}
  {\bibinfo  {journal} {Phys. Rev. B}\ }\textbf {\bibinfo {volume} {107}},\
  \bibinfo {pages} {L241405} (\bibinfo {year}
  {2023}{\natexlab{b}})}\BibitemShut {NoStop}%
\bibitem [{\citenamefont {Pekola}\ and\ \citenamefont
  {Karimi}(2021)}]{PekolaPRM21}%
  \BibitemOpen
  \bibfield  {author} {\bibinfo {author} {\bibfnamefont {J.~P.}\ \bibnamefont
  {Pekola}}\ and\ \bibinfo {author} {\bibfnamefont {B.}~\bibnamefont
  {Karimi}},\ }\bibfield  {title} {\enquote {\bibinfo {title} {Colloquium:
  Quantum heat transport in condensed matter systems},}\ }\href {\doibase
  10.1103/RevModPhys.93.041001} {\bibfield  {journal} {\bibinfo  {journal}
  {Rev. Mod. Phys.}\ }\textbf {\bibinfo {volume} {93}},\ \bibinfo {pages}
  {041001} (\bibinfo {year} {2021})}\BibitemShut {NoStop}%
\bibitem [{\citenamefont {L\"u}\ \emph {et~al.}(2016)\citenamefont {L\"u},
  \citenamefont {Wang}, \citenamefont {Hedeg\aa{}rd},\ and\ \citenamefont
  {Brandbyge}}]{LvPRB16}%
  \BibitemOpen
  \bibfield  {author} {\bibinfo {author} {\bibfnamefont {J.-T.}\ \bibnamefont
  {L\"u}}, \bibinfo {author} {\bibfnamefont {J.-S.}\ \bibnamefont {Wang}},
  \bibinfo {author} {\bibfnamefont {P.}~\bibnamefont {Hedeg\aa{}rd}}, \ and\
  \bibinfo {author} {\bibfnamefont {M.}~\bibnamefont {Brandbyge}},\ }\bibfield
  {title} {\enquote {\bibinfo {title} {Electron and phonon drag in
  thermoelectric transport through coherent molecular conductors},}\ }\href
  {\doibase 10.1103/PhysRevB.93.205404} {\bibfield  {journal} {\bibinfo
  {journal} {Phys. Rev. B}\ }\textbf {\bibinfo {volume} {93}},\ \bibinfo
  {pages} {205404} (\bibinfo {year} {2016})}\BibitemShut {NoStop}%
\bibitem [{\citenamefont {Li}\ \emph {et~al.}(2021)\citenamefont {Li},
  \citenamefont {Li}, \citenamefont {Han}, \citenamefont {Zheng}, \citenamefont
  {Li}, \citenamefont {Li}, \citenamefont {Fan},\ and\ \citenamefont
  {Qiu}}]{Li21NRM}%
  \BibitemOpen
  \bibfield  {author} {\bibinfo {author} {\bibfnamefont {Y.}~\bibnamefont
  {Li}}, \bibinfo {author} {\bibfnamefont {W.}~\bibnamefont {Li}}, \bibinfo
  {author} {\bibfnamefont {T.}~\bibnamefont {Han}}, \bibinfo {author}
  {\bibfnamefont {X.}~\bibnamefont {Zheng}}, \bibinfo {author} {\bibfnamefont
  {J.}~\bibnamefont {Li}}, \bibinfo {author} {\bibfnamefont {B.}~\bibnamefont
  {Li}}, \bibinfo {author} {\bibfnamefont {S.}~\bibnamefont {Fan}}, \ and\
  \bibinfo {author} {\bibfnamefont {C.-W.}\ \bibnamefont {Qiu}},\ }\bibfield
  {title} {\enquote {\bibinfo {title} {Transforming heat transfer with thermal
  metamaterials and devices},}\ }\href {\doibase 10.1038/s41578-021-00283-2}
  {\bibfield  {journal} {\bibinfo  {journal} {Nature Reviews Materials}\
  }\textbf {\bibinfo {volume} {6}},\ \bibinfo {pages} {488--507} (\bibinfo
  {year} {2021})}\BibitemShut {NoStop}%
\bibitem [{\citenamefont {Nian}\ \emph
  {et~al.}(2023{\natexlab{c}})\citenamefont {Nian}, \citenamefont {Hu},
  \citenamefont {Xiong}, \citenamefont {L\"u},\ and\ \citenamefont
  {Zheng}}]{NianPRB23}%
  \BibitemOpen
  \bibfield  {author} {\bibinfo {author} {\bibfnamefont {L.-L.}\ \bibnamefont
  {Nian}}, \bibinfo {author} {\bibfnamefont {S.}~\bibnamefont {Hu}}, \bibinfo
  {author} {\bibfnamefont {L.}~\bibnamefont {Xiong}}, \bibinfo {author}
  {\bibfnamefont {J.-T.}\ \bibnamefont {L\"u}}, \ and\ \bibinfo {author}
  {\bibfnamefont {B.}~\bibnamefont {Zheng}},\ }\bibfield  {title} {\enquote
  {\bibinfo {title} {Photon-assisted electron transport across a quantum phase
  transition},}\ }\href {\doibase 10.1103/PhysRevB.108.085430} {\bibfield
  {journal} {\bibinfo  {journal} {Phys. Rev. B}\ }\textbf {\bibinfo {volume}
  {108}},\ \bibinfo {pages} {085430} (\bibinfo {year}
  {2023}{\natexlab{c}})}\BibitemShut {NoStop}%
\bibitem [{\citenamefont {Jiang}\ \emph {et~al.}(2015)\citenamefont {Jiang},
  \citenamefont {Kulkarni}, \citenamefont {Segal},\ and\ \citenamefont
  {Imry}}]{Jiangtransistors}%
  \BibitemOpen
  \bibfield  {author} {\bibinfo {author} {\bibfnamefont {J.-H.}\ \bibnamefont
  {Jiang}}, \bibinfo {author} {\bibfnamefont {M.}~\bibnamefont {Kulkarni}},
  \bibinfo {author} {\bibfnamefont {D.}~\bibnamefont {Segal}}, \ and\ \bibinfo
  {author} {\bibfnamefont {Y.}~\bibnamefont {Imry}},\ }\bibfield  {title}
  {\enquote {\bibinfo {title} {Phonon thermoelectric transistors and
  rectifiers},}\ }\href {\doibase 10.1103/PhysRevB.92.045309} {\bibfield
  {journal} {\bibinfo  {journal} {Phys. Rev. B}\ }\textbf {\bibinfo {volume}
  {92}},\ \bibinfo {pages} {045309} (\bibinfo {year} {2015})}\BibitemShut
  {NoStop}%
\bibitem [{\citenamefont {Lu}\ \emph {et~al.}(2020)\citenamefont {Lu},
  \citenamefont {Wang}, \citenamefont {Wang},\ and\ \citenamefont
  {Jiang}}]{MyPRBtransistor}%
  \BibitemOpen
  \bibfield  {author} {\bibinfo {author} {\bibfnamefont {J.}~\bibnamefont
  {Lu}}, \bibinfo {author} {\bibfnamefont {R.}~\bibnamefont {Wang}}, \bibinfo
  {author} {\bibfnamefont {C.}~\bibnamefont {Wang}}, \ and\ \bibinfo {author}
  {\bibfnamefont {J.-H.}\ \bibnamefont {Jiang}},\ }\bibfield  {title} {\enquote
  {\bibinfo {title} {Brownian thermal transistors and refrigerators in
  mesoscopic systems},}\ }\href {\doibase 10.1103/PhysRevB.102.125405}
  {\bibfield  {journal} {\bibinfo  {journal} {Phys. Rev. B}\ }\textbf {\bibinfo
  {volume} {102}},\ \bibinfo {pages} {125405} (\bibinfo {year}
  {2020})}\BibitemShut {NoStop}%
\bibitem [{\citenamefont {Lu}\ \emph {et~al.}(2019)\citenamefont {Lu},
  \citenamefont {Wang}, \citenamefont {Ren}, \citenamefont {Kulkarni},\ and\
  \citenamefont {Jiang}}]{MyPRBdiode}%
  \BibitemOpen
  \bibfield  {author} {\bibinfo {author} {\bibfnamefont {J.}~\bibnamefont
  {Lu}}, \bibinfo {author} {\bibfnamefont {R.}~\bibnamefont {Wang}}, \bibinfo
  {author} {\bibfnamefont {J.}~\bibnamefont {Ren}}, \bibinfo {author}
  {\bibfnamefont {M.}~\bibnamefont {Kulkarni}}, \ and\ \bibinfo {author}
  {\bibfnamefont {J.-H.}\ \bibnamefont {Jiang}},\ }\bibfield  {title} {\enquote
  {\bibinfo {title} {Quantum-dot circuit-qed thermoelectric diodes and
  transistors},}\ }\href {\doibase 10.1103/PhysRevB.99.035129} {\bibfield
  {journal} {\bibinfo  {journal} {Phys. Rev. B}\ }\textbf {\bibinfo {volume}
  {99}},\ \bibinfo {pages} {035129} (\bibinfo {year} {2019})}\BibitemShut
  {NoStop}%
\bibitem [{\citenamefont {Campisi}\ \emph {et~al.}(2011)\citenamefont
  {Campisi}, \citenamefont {H\"anggi},\ and\ \citenamefont
  {Talkner}}]{CampisiRMP}%
  \BibitemOpen
  \bibfield  {author} {\bibinfo {author} {\bibfnamefont {M.}~\bibnamefont
  {Campisi}}, \bibinfo {author} {\bibfnamefont {P.}~\bibnamefont {H\"anggi}}, \
  and\ \bibinfo {author} {\bibfnamefont {P.}~\bibnamefont {Talkner}},\
  }\bibfield  {title} {\enquote {\bibinfo {title} {Colloquium: Quantum
  fluctuation relations: Foundations and applications},}\ }\href {\doibase
  10.1103/RevModPhys.83.771} {\bibfield  {journal} {\bibinfo  {journal} {Rev.
  Mod. Phys.}\ }\textbf {\bibinfo {volume} {83}},\ \bibinfo {pages} {771--791}
  (\bibinfo {year} {2011})}\BibitemShut {NoStop}%
\bibitem [{Sup()}]{Supp}%
  \BibitemOpen
  \href@noop {} {\bibinfo  {journal} {See supplementary materials for the
  detailed derivations.}\ }\BibitemShut {NoStop}%
\bibitem [{\citenamefont {Wang}\ \emph
  {et~al.}(2022{\natexlab{b}})\citenamefont {Wang}, \citenamefont {Wang},
  \citenamefont {Chen}, \citenamefont {Wang},\ and\ \citenamefont
  {Ren}}]{wangpump}%
  \BibitemOpen
\bibfield  {journal} {  }\bibfield  {author} {\bibinfo {author} {\bibfnamefont
  {Z.}~\bibnamefont {Wang}}, \bibinfo {author} {\bibfnamefont {L.}~\bibnamefont
  {Wang}}, \bibinfo {author} {\bibfnamefont {J.}~\bibnamefont {Chen}}, \bibinfo
  {author} {\bibfnamefont {C.}~\bibnamefont {Wang}}, \ and\ \bibinfo {author}
  {\bibfnamefont {J.}~\bibnamefont {Ren}},\ }\bibfield  {title} {\enquote
  {\bibinfo {title} {Geometric heat pump: Controlling thermal transport with
  time-dependent modulations},}\ }\href {\doibase 10.1007/s11467-021-1095-4}
  {\bibfield  {journal} {\bibinfo  {journal} {Front. Phys.}\ }\textbf {\bibinfo
  {volume} {17}},\ \bibinfo {pages} {1--14} (\bibinfo {year}
  {2022}{\natexlab{b}})}\BibitemShut {NoStop}%
\bibitem [{\citenamefont {Yamamoto}\ and\ \citenamefont
  {Hatano}(2015)}]{YamamotoPRE15}%
  \BibitemOpen
  \bibfield  {author} {\bibinfo {author} {\bibfnamefont {K.}~\bibnamefont
  {Yamamoto}}\ and\ \bibinfo {author} {\bibfnamefont {N.}~\bibnamefont
  {Hatano}},\ }\bibfield  {title} {\enquote {\bibinfo {title} {Thermodynamics
  of the mesoscopic thermoelectric heat engine beyond the linear-response
  regime},}\ }\href {\doibase 10.1103/PhysRevE.92.042165} {\bibfield  {journal}
  {\bibinfo  {journal} {Phys. Rev. E}\ }\textbf {\bibinfo {volume} {92}},\
  \bibinfo {pages} {042165} (\bibinfo {year} {2015})}\BibitemShut {NoStop}%
\bibitem [{\citenamefont {Jiang}\ and\ \citenamefont {Imry}(2016)}]{JiangCRP}%
  \BibitemOpen
  \bibfield  {author} {\bibinfo {author} {\bibfnamefont {J.-H.}\ \bibnamefont
  {Jiang}}\ and\ \bibinfo {author} {\bibfnamefont {Y.}~\bibnamefont {Imry}},\
  }\bibfield  {title} {\enquote {\bibinfo {title} {Linear and nonlinear
  mesoscopic thermoelectric transport with coupling with heat baths},}\ }\href
  {\doibase https://doi.org/10.1016/j.crhy.2016.08.006} {\bibfield  {journal}
  {\bibinfo  {journal} {C. R. Phys.}\ }\textbf {\bibinfo {volume} {17}},\
  \bibinfo {pages} {1047 -- 1059} (\bibinfo {year} {2016})}\BibitemShut
  {NoStop}%
\bibitem [{\citenamefont {Lu}\ \emph {et~al.}(2024)\citenamefont {Lu},
  \citenamefont {Wang}, \citenamefont {Ren}, \citenamefont {Wang},\ and\
  \citenamefont {Jiang}}]{Lu23Coh}%
  \BibitemOpen
  \bibfield  {author} {\bibinfo {author} {\bibfnamefont {Jincheng}\
  \bibnamefont {Lu}}, \bibinfo {author} {\bibfnamefont {Zi}~\bibnamefont
  {Wang}}, \bibinfo {author} {\bibfnamefont {Jie}\ \bibnamefont {Ren}},
  \bibinfo {author} {\bibfnamefont {Chen}\ \bibnamefont {Wang}}, \ and\
  \bibinfo {author} {\bibfnamefont {Jian-Hua}\ \bibnamefont {Jiang}},\
  }\bibfield  {title} {\enquote {\bibinfo {title} {Coherence-enhanced
  thermodynamic performance in a periodically driven inelastic heat engine},}\
  }\href {\doibase 10.1103/PhysRevB.109.125407} {\bibfield  {journal} {\bibinfo
   {journal} {Phys. Rev. B}\ }\textbf {\bibinfo {volume} {109}},\ \bibinfo
  {pages} {125407} (\bibinfo {year} {2024})}\BibitemShut {NoStop}%
\bibitem [{\citenamefont {Zhang}\ \emph {et~al.}(2010)\citenamefont {Zhang},
  \citenamefont {Wang},\ and\ \citenamefont {Li}}]{ZhangPRB10}%
  \BibitemOpen
  \bibfield  {author} {\bibinfo {author} {\bibfnamefont {L.}~\bibnamefont
  {Zhang}}, \bibinfo {author} {\bibfnamefont {J.-S.}\ \bibnamefont {Wang}}, \
  and\ \bibinfo {author} {\bibfnamefont {B.}~\bibnamefont {Li}},\ }\bibfield
  {title} {\enquote {\bibinfo {title} {Ballistic thermal rectification in
  nanoscale three-terminal junctions},}\ }\href {\doibase
  10.1103/PhysRevB.81.100301} {\bibfield  {journal} {\bibinfo  {journal} {Phys.
  Rev. B}\ }\textbf {\bibinfo {volume} {81}},\ \bibinfo {pages} {100301}
  (\bibinfo {year} {2010})}\BibitemShut {NoStop}%
\bibitem [{\citenamefont {Ren}\ and\ \citenamefont {Zhu}(2013)}]{RenPRB13}%
  \BibitemOpen
  \bibfield  {author} {\bibinfo {author} {\bibfnamefont {J.}~\bibnamefont
  {Ren}}\ and\ \bibinfo {author} {\bibfnamefont {J.-X.}\ \bibnamefont {Zhu}},\
  }\bibfield  {title} {\enquote {\bibinfo {title} {Theory of asymmetric and
  negative differential magnon tunneling under temperature bias: Towards a spin
  seebeck diode and transistor},}\ }\href {\doibase 10.1103/PhysRevB.88.094427}
  {\bibfield  {journal} {\bibinfo  {journal} {Phys. Rev. B}\ }\textbf {\bibinfo
  {volume} {88}},\ \bibinfo {pages} {094427} (\bibinfo {year}
  {2013})}\BibitemShut {NoStop}%
\bibitem [{\citenamefont {Yang}\ \emph {et~al.}(2018)\citenamefont {Yang},
  \citenamefont {Chen}, \citenamefont {Wang}, \citenamefont {Li},\ and\
  \citenamefont {Zhang}}]{ZhangPRE18}%
  \BibitemOpen
  \bibfield  {author} {\bibinfo {author} {\bibfnamefont {Y.}~\bibnamefont
  {Yang}}, \bibinfo {author} {\bibfnamefont {H.}~\bibnamefont {Chen}}, \bibinfo
  {author} {\bibfnamefont {H.}~\bibnamefont {Wang}}, \bibinfo {author}
  {\bibfnamefont {N.}~\bibnamefont {Li}}, \ and\ \bibinfo {author}
  {\bibfnamefont {L.}~\bibnamefont {Zhang}},\ }\bibfield  {title} {\enquote
  {\bibinfo {title} {Optimal thermal rectification of heterojunctions under
  fourier law},}\ }\href {\doibase 10.1103/PhysRevE.98.042131} {\bibfield
  {journal} {\bibinfo  {journal} {Phys. Rev. E}\ }\textbf {\bibinfo {volume}
  {98}},\ \bibinfo {pages} {042131} (\bibinfo {year} {2018})}\BibitemShut
  {NoStop}%
\bibitem [{\citenamefont {Zhang}\ and\ \citenamefont
  {Su}(2021)}]{ZhangPhysicaA}%
  \BibitemOpen
  \bibfield  {author} {\bibinfo {author} {\bibfnamefont {Y.}~\bibnamefont
  {Zhang}}\ and\ \bibinfo {author} {\bibfnamefont {S.}~\bibnamefont {Su}},\
  }\bibfield  {title} {\enquote {\bibinfo {title} {Thermal rectification and
  negative differential thermal conductance based on a parallel-coupled double
  quantum-dot},}\ }\href {\doibase https://doi.org/10.1016/j.physa.2021.126347}
  {\bibfield  {journal} {\bibinfo  {journal} {Physica A: Statistical Mechanics
  and its Applications}\ }\textbf {\bibinfo {volume} {584}},\ \bibinfo {pages}
  {126347} (\bibinfo {year} {2021})}\BibitemShut {NoStop}%
\bibitem [{\citenamefont {Wang}\ \emph {et~al.}(2019)\citenamefont {Wang},
  \citenamefont {Xu}, \citenamefont {Liu},\ and\ \citenamefont
  {Gao}}]{WangPRE19}%
  \BibitemOpen
  \bibfield  {author} {\bibinfo {author} {\bibfnamefont {C.}~\bibnamefont
  {Wang}}, \bibinfo {author} {\bibfnamefont {D.}~\bibnamefont {Xu}}, \bibinfo
  {author} {\bibfnamefont {H.}~\bibnamefont {Liu}}, \ and\ \bibinfo {author}
  {\bibfnamefont {X.}~\bibnamefont {Gao}},\ }\bibfield  {title} {\enquote
  {\bibinfo {title} {Thermal rectification and heat amplification in a
  nonequilibrium v-type three-level system},}\ }\href {\doibase
  10.1103/PhysRevE.99.042102} {\bibfield  {journal} {\bibinfo  {journal} {Phys.
  Rev. E}\ }\textbf {\bibinfo {volume} {99}},\ \bibinfo {pages} {042102}
  (\bibinfo {year} {2019})}\BibitemShut {NoStop}%
\bibitem [{\citenamefont {Diaz}\ and\ \citenamefont
  {Sanchez}(2021)}]{DazNJP21}%
  \BibitemOpen
  \bibfield  {author} {\bibinfo {author} {\bibfnamefont {I.}~\bibnamefont
  {Diaz}}\ and\ \bibinfo {author} {\bibfnamefont {R.}~\bibnamefont {Sanchez}},\
  }\bibfield  {title} {\enquote {\bibinfo {title} {The qutrit as a heat diode
  and circulator},}\ }\href {\doibase 10.1088/1367-2630/ac4211} {\bibfield
  {journal} {\bibinfo  {journal} {New J. Phys.}\ }\textbf {\bibinfo {volume}
  {23}},\ \bibinfo {pages} {125006} (\bibinfo {year} {2021})}\BibitemShut
  {NoStop}%
\bibitem [{\citenamefont {Tesser}\ \emph {et~al.}(2022)\citenamefont {Tesser},
  \citenamefont {Bhandari}, \citenamefont {Erdman}, \citenamefont {Paladino},
  \citenamefont {Fazio},\ and\ \citenamefont {Taddei}}]{TesserNJP22}%
  \BibitemOpen
  \bibfield  {author} {\bibinfo {author} {\bibfnamefont {L.}~\bibnamefont
  {Tesser}}, \bibinfo {author} {\bibfnamefont {B.}~\bibnamefont {Bhandari}},
  \bibinfo {author} {\bibfnamefont {P.~Andrea}\ \bibnamefont {Erdman}},
  \bibinfo {author} {\bibfnamefont {E.}~\bibnamefont {Paladino}}, \bibinfo
  {author} {\bibfnamefont {R.}~\bibnamefont {Fazio}}, \ and\ \bibinfo {author}
  {\bibfnamefont {F.}~\bibnamefont {Taddei}},\ }\bibfield  {title} {\enquote
  {\bibinfo {title} {Heat rectification through single and coupled quantum
  dots},}\ }\href {\doibase 10.1088/1367-2630/ac53b8} {\bibfield  {journal}
  {\bibinfo  {journal} {New J. Phys.}\ }\textbf {\bibinfo {volume} {24}},\
  \bibinfo {pages} {035001} (\bibinfo {year} {2022})}\BibitemShut {NoStop}%
\bibitem [{\citenamefont {Khandelwal}\ \emph {et~al.}(2023)\citenamefont
  {Khandelwal}, \citenamefont {Perarnau-Llobet}, \citenamefont {Seah},
  \citenamefont {Brunner},\ and\ \citenamefont {Haack}}]{KhandelwalPRR23}%
  \BibitemOpen
  \bibfield  {author} {\bibinfo {author} {\bibfnamefont {S.}~\bibnamefont
  {Khandelwal}}, \bibinfo {author} {\bibfnamefont {M.}~\bibnamefont
  {Perarnau-Llobet}}, \bibinfo {author} {\bibfnamefont {S.}~\bibnamefont
  {Seah}}, \bibinfo {author} {\bibfnamefont {N.}~\bibnamefont {Brunner}}, \
  and\ \bibinfo {author} {\bibfnamefont {G.}~\bibnamefont {Haack}},\ }\bibfield
   {title} {\enquote {\bibinfo {title} {Characterizing the performance of heat
  rectifiers},}\ }\href {\doibase 10.1103/PhysRevResearch.5.013129} {\bibfield
  {journal} {\bibinfo  {journal} {Phys. Rev. Res.}\ }\textbf {\bibinfo {volume}
  {5}},\ \bibinfo {pages} {013129} (\bibinfo {year} {2023})}\BibitemShut
  {NoStop}%
\bibitem [{\citenamefont {Joulain}\ \emph {et~al.}(2016)\citenamefont
  {Joulain}, \citenamefont {Drevillon}, \citenamefont {Ezzahri},\ and\
  \citenamefont {Ordonez-Miranda}}]{Transistor9}%
  \BibitemOpen
  \bibfield  {author} {\bibinfo {author} {\bibfnamefont {K.}~\bibnamefont
  {Joulain}}, \bibinfo {author} {\bibfnamefont {J.}~\bibnamefont {Drevillon}},
  \bibinfo {author} {\bibfnamefont {Y.}~\bibnamefont {Ezzahri}}, \ and\
  \bibinfo {author} {\bibfnamefont {J.}~\bibnamefont {Ordonez-Miranda}},\
  }\bibfield  {title} {\enquote {\bibinfo {title} {Quantum thermal
  transistor},}\ }\href {\doibase 10.1103/PhysRevLett.116.200601} {\bibfield
  {journal} {\bibinfo  {journal} {Phys. Rev. Lett.}\ }\textbf {\bibinfo
  {volume} {116}},\ \bibinfo {pages} {200601} (\bibinfo {year}
  {2016})}\BibitemShut {NoStop}%
\bibitem [{\citenamefont {Guo}\ \emph {et~al.}(2018)\citenamefont {Guo},
  \citenamefont {Liu},\ and\ \citenamefont {Yu}}]{YuPRE18}%
  \BibitemOpen
  \bibfield  {author} {\bibinfo {author} {\bibfnamefont {B.-q.}\ \bibnamefont
  {Guo}}, \bibinfo {author} {\bibfnamefont {T.}~\bibnamefont {Liu}}, \ and\
  \bibinfo {author} {\bibfnamefont {C.-s.}\ \bibnamefont {Yu}},\ }\bibfield
  {title} {\enquote {\bibinfo {title} {Quantum thermal transistor based on
  qubit-qutrit coupling},}\ }\href {\doibase 10.1103/PhysRevE.98.022118}
  {\bibfield  {journal} {\bibinfo  {journal} {Phys. Rev. E}\ }\textbf {\bibinfo
  {volume} {98}},\ \bibinfo {pages} {022118} (\bibinfo {year}
  {2018})}\BibitemShut {NoStop}%
\bibitem [{\citenamefont {Yang}\ \emph {et~al.}(2023)\citenamefont {Yang},
  \citenamefont {Zhao},\ and\ \citenamefont {Zhang}}]{ZhangAPL23}%
  \BibitemOpen
  \bibfield  {author} {\bibinfo {author} {\bibfnamefont {Y.}~\bibnamefont
  {Yang}}, \bibinfo {author} {\bibfnamefont {Y.}~\bibnamefont {Zhao}}, \ and\
  \bibinfo {author} {\bibfnamefont {L.}~\bibnamefont {Zhang}},\ }\bibfield
  {title} {\enquote {\bibinfo {title} {{A high-performance thermal transistor
  based on interfacial negative differential thermal resistance}},}\ }\href
  {\doibase 10.1063/5.0149544} {\bibfield  {journal} {\bibinfo  {journal}
  {Applied Physics Letters}\ }\textbf {\bibinfo {volume} {122}},\ \bibinfo
  {pages} {232201} (\bibinfo {year} {2023})}\BibitemShut {NoStop}%
\bibitem [{\citenamefont {Li}\ \emph {et~al.}(2004)\citenamefont {Li},
  \citenamefont {Wang},\ and\ \citenamefont {Casati}}]{LiPRL04}%
  \BibitemOpen
  \bibfield  {author} {\bibinfo {author} {\bibfnamefont {B.}~\bibnamefont
  {Li}}, \bibinfo {author} {\bibfnamefont {L.}~\bibnamefont {Wang}}, \ and\
  \bibinfo {author} {\bibfnamefont {G.}~\bibnamefont {Casati}},\ }\bibfield
  {title} {\enquote {\bibinfo {title} {Thermal diode: Rectification of heat
  flux},}\ }\href {\doibase 10.1103/PhysRevLett.93.184301} {\bibfield
  {journal} {\bibinfo  {journal} {Phys. Rev. Lett.}\ }\textbf {\bibinfo
  {volume} {93}},\ \bibinfo {pages} {184301} (\bibinfo {year}
  {2004})}\BibitemShut {NoStop}%
\bibitem [{\citenamefont {Li}\ \emph {et~al.}(2006)\citenamefont {Li},
  \citenamefont {Wang},\ and\ \citenamefont {Casati}}]{LiAPL06}%
  \BibitemOpen
  \bibfield  {author} {\bibinfo {author} {\bibfnamefont {B.}~\bibnamefont
  {Li}}, \bibinfo {author} {\bibfnamefont {L.}~\bibnamefont {Wang}}, \ and\
  \bibinfo {author} {\bibfnamefont {G.}~\bibnamefont {Casati}},\ }\bibfield
  {title} {\enquote {\bibinfo {title} {{Negative differential thermal
  resistance and thermal transistor}},}\ }\href {\doibase 10.1063/1.2191730}
  {\bibfield  {journal} {\bibinfo  {journal} {Applied Physics Letters}\
  }\textbf {\bibinfo {volume} {88}},\ \bibinfo {pages} {143501} (\bibinfo
  {year} {2006})}\BibitemShut {NoStop}%
\bibitem [{\citenamefont {S\'anchez}\ \emph {et~al.}(2017)\citenamefont
  {S\'anchez}, \citenamefont {Thierschmann},\ and\ \citenamefont
  {Molenkamp}}]{Transistor1}%
  \BibitemOpen
  \bibfield  {author} {\bibinfo {author} {\bibfnamefont {R.}~\bibnamefont
  {S\'anchez}}, \bibinfo {author} {\bibfnamefont {H.}~\bibnamefont
  {Thierschmann}}, \ and\ \bibinfo {author} {\bibfnamefont {L.~W.}\
  \bibnamefont {Molenkamp}},\ }\bibfield  {title} {\enquote {\bibinfo {title}
  {All-thermal transistor based on stochastic switching},}\ }\href {\doibase
  10.1103/PhysRevB.95.241401} {\bibfield  {journal} {\bibinfo  {journal} {Phys.
  Rev. B}\ }\textbf {\bibinfo {volume} {95}},\ \bibinfo {pages} {241401}
  (\bibinfo {year} {2017})}\BibitemShut {NoStop}%
\bibitem [{\citenamefont {Wang}\ \emph {et~al.}(2018)\citenamefont {Wang},
  \citenamefont {Chen}, \citenamefont {Sun},\ and\ \citenamefont
  {Ren}}]{WangPRA18}%
  \BibitemOpen
  \bibfield  {author} {\bibinfo {author} {\bibfnamefont {C.}~\bibnamefont
  {Wang}}, \bibinfo {author} {\bibfnamefont {X.-M.}\ \bibnamefont {Chen}},
  \bibinfo {author} {\bibfnamefont {K.-W.}\ \bibnamefont {Sun}}, \ and\
  \bibinfo {author} {\bibfnamefont {J.}~\bibnamefont {Ren}},\ }\bibfield
  {title} {\enquote {\bibinfo {title} {Heat amplification and negative
  differential thermal conductance in a strongly coupled nonequilibrium
  spin-boson system},}\ }\href {\doibase 10.1103/PhysRevA.97.052112} {\bibfield
   {journal} {\bibinfo  {journal} {Phys. Rev. A}\ }\textbf {\bibinfo {volume}
  {97}},\ \bibinfo {pages} {052112} (\bibinfo {year} {2018})}\BibitemShut
  {NoStop}%
\end{thebibliography}%

\end{document}